\documentstyle[12pt]{article}

\begin{document}
%\preprint{}

\title{Hausdorff dimension of repellors in low sensitive systems}

\author{ {\it A. E. Motter$^a$ and P. S. Letelier$^b$}\\
Departamento de Matem\'atica Aplicada - IMECC \\
Universidade Estadual de Campinas, Unicamp\\
13081-970 Campinas, SP, Brazil}

\date{}
\maketitle

\begin{abstract}

Methods to estimate the Hausdorff dimension of
invariant sets of scattering systems are presented.
Based on the levels' hierarchical structure of the
time delay function, these techniques can be used
in systems whose future-invariant-set
codimensions are approximately equal to or greater than one.
The discussion is illustrated by a numerical example
of a scatterer built with four hard spheres located at
the vertices of a regular tetrahedron.

\end{abstract}

\vskip 1.0truecm

PACS numbers: 05.45.Df

Keywords: chaotic scattering, fractal, Hausdorff dimension, repellor. 

$^a$E-mail: motter@ime.unicamp.br.

$^b$E-mail: letelier@ime.unicamp.br.

\newpage

Usually, in a scattering process almost all particles outcome
asymptotically free after interacting with a potential.
Some particles, however, remain in the
scattering region bouncing forever under the action of
the potential. The nature of these trapped orbits is
in general very intricate, including fractal dimensions
and chaotic behaviour.
The dimension seems to be the most fundamental
quantity characterising this set. When the dimension
is large enough, scattered particles may present sensitive
dependence on initial conditions reminiscent of the
chaotic dynamics of the trapped orbits. This phenomenon
allows the computation of the {\it uncertainty dimension}
of the bounded orbits, defined as follows.
For a scattering system, colour by black initial conditions
of particles scattered, for instance, to the right-hand side
of the scattering centre and by white initial
conditions corresponding to particles scattered to the
left-hand side. An initial condition of a given colour at a
distance less than $\varepsilon$ of initial conditions
of a different colour is labelled $\varepsilon$-uncertain.
The final state (left or right)
of a particle associated to an
$\varepsilon$-uncertain initial condition is uncertain
if this point is determined only within a precision of
order $\varepsilon$. In the limit of small $\varepsilon$,
the fraction of such $\varepsilon$-uncertain points is
$f(\varepsilon)\sim \varepsilon^{N-D_u}$, for $N$ denoting the
dimension of the subspace where the initial conditions are
taken, and $D_u$ the corresponding uncertainty dimension of the
intersection of the trapped orbits with this subspace.
Therefore $D_u$ can be computed by fitting $\ln f(\varepsilon)$
as a function of $\ln(\varepsilon)$. This procedure works in
systems where the set of bounded orbits has dimension
greater than $N-1$ in the $N$-dimensional
subspace of initial conditions.
That is supported by the fact that in this case
$D_u$ is expected to be equal to a more
general concept of dimension, the {\it Hausdorff dimension}.

The Hausdorff dimension of a set $A$ is defined as follows
\cite{ott}: Let $S_j$ denote a countable collection of subsets
of the Euclidean space such that the diameters $\epsilon_j$
of the $S_j$ are all less than or equal to $\delta$, and
such that the $S_j$ are a covering of $A$. We define the
Hausdorff sum
\begin{equation}
K_{\delta}(d)=\inf_{S_j}\sum_{j=1}(\epsilon_j)^d ,
\label{1}
\end{equation}
where the infimum is taken on the set of all the possible
countable coverings of $A$ satisfying $\epsilon_j\leq\delta$.
Then we define the $d$-dimensional Hausdorff measure
\begin{equation}
K(d)=\lim_{\delta\rightarrow 0}K_{\delta}(d).
\label{2}
\end{equation}
It can be shown that $K(d)$ is infinite if $d$ is less than
some critical value $D_H$, and is zero if $d$ is greater
than $D_H$. This critical value is the Hausdorff dimension
of the set $A$.

Among the several existing concepts of dimension,
the Hausdorff dimension is the one that better
implement the intuitive notion of dimension that
we have in mind. Unfortunately, the uncertainty
methods cannot be used to estimate $D_H$
when its value is less than or equal to $N-1$.
Even when $D_H$ is slightly greater than $N-1$,
its evaluation through the numerical computation
of the uncertainty dimension is highly inaccurate.
The aim of these work is to present
methods to compute the Hausdorff dimension of
the set of trapped orbits - the future invariant set ($I_f$) -
when uncertainty
methods do not apply. The determination of this
dimension is important, for instance, to understand
the onset of chaos in scattering systems when a
system parameter is varied.

First in this communication,
we introduce a method to take into account the
infimum requirement appearing in the definition of
the Hausdorff dimension,
that allows the direct computation of $D_H(I_f)$.
Next, the Hausdorff dimension is estimated
from the box-counting dimension computed by
using a proper-interior-maximum-{\it like} procedure.
Finally, we define a new dimension,
the {\it self-similarity dimension}, which is also
used to approximate $D_H(I_f)$.
Our methods can be applied for $D_H(I_f)>N-1$ as well as
for $D_H(I_f)\leq N-1$.
The discussion is illustrated along the text
by a numerical example
of a simple scattering system presenting 
$D_H(I_f)\leq N-1$ and $D_H(I_f)>N-1$
for two ranges of values of its control parameter.

The example we shall consider is due to
Chen {\it et al.} \cite{chen},
who investigated a system consisting of a particle
bouncing between four identical hard
spheres placed at the vertices
of a regular tetrahedron of unit edge.
The centres of the spheres are located at the
coordinates
$(0,$ $0,$ $\sqrt{2}/3)$,
$(1/2,$ $-1/2\sqrt{3},$ $0)$,
$(-1/2,$ $-1/2\sqrt{3},$ $0)$
and $(0,$ $1/\sqrt{3},$ $0)$,
and their radius $R$ is the
variable parameter of the system.
These authors contributed with a procedure to directly
evaluate the Hausdorff dimension $\tilde{D}_H(R)$ on 1-dimensional
lines. By taking lines on the plane
$P=\{z=-k$ $(k>R),p_x=p_y=0,p_z=1\}$,
they computed the Hausdorff dimension $D_H(R)$ of the
intersection of $I_f$ with $P$,
obtaining $D_H(0.48)=1+\tilde{D}_H(0.48)=1.4$.
It is reasonable to presume that $D_H$ is an increasing
function of $R$. Here we are interested in computing
$D_H$ for small values of the radius,
where the dimension of $I_f\cap P$ is expected
to be less than one (codimension greater than one).

We consider initial conditions $(x_0,y_0)$ on $P$,
and we define the time delay function $T(x_0,y_0)$
as the number of collisions of the particle
with the spheres. We also introduce
$C_n=\{ (x_0,y_0)\mid T(x_0,y_0)\geq n\}$.
$C_1$ is the union of 4 discs, the projection on $P$
of the four spheres. $C_2$ is the union of $4\cdot 3=12$
deformed discs, the projection on $P$ of the reflection
on each sphere of the other three spheres.
$C_3$ is the union of $4\cdot 3^2=36$ deformed discs, and so on.
Therefore, for sufficiently small $R$,
$C_n$ is the union of $4\cdot 3^{n-1}$ deformed discs.
The construction of the Cantor structure of the set
$I_f\cap P=\cap_{n=1}^{\infty}C_n$ can be followed in Fig. 1,
where we show $C_1$, $C_2$, $C_3$ and $C_4$ for $R=0.37$.

Labelling the spheres by 1, 2, 3 and 4, the left shift
symbolic dynamics of the future invariant set consists
of sequences $s_1s_2s_3...$ of
1, 2, 3 and 4, where $s_n$ is the number of the
sphere of the $n^{th}$ collision. The particle cannot
collide two successive times with the same sphere,
and hence the only constraint is
$s_{n+1}\neq s_n$ (at least for $R$ small enough).
A relevant point here is that each one of the
$4\cdot 3^{n-1}$ deformed discs of $C_n$ can be identified
by a sequence of collisions,
$c_{n,j}=[s_1s_2...s_n]$
for $j=1,2,...,4\cdot 3^{n-1}$. It allows us to define
the Hausdorff sum
\begin{equation}
K_n(d)=\sum_{j=1}^{4\cdot 3^{n-1}}|c_{n,j}|^d,
\label{sum}
\end{equation}
where $|c_{n,j}|$ is the diameter of $c_{n,j}$,
$|c_{n,j}|=\sup_{\vec{x},\vec{y}\in c_{n,j}}|\vec{x}-\vec{y}|$,
which can be easily estimated from a sample of
initial conditions. 
The $d-$dimensional Hausdorff measure
$K(d)$ is supposed to be the limit of $K_n(d)$
for $n\rightarrow\infty$. $K(d)$ is infinite
for $d$ less than the Hausdorff
dimension $D_H$, and is zero for $d$ greater than $D_H$
\cite{ott}.
Since it is not known how to compute
lim$_{n\rightarrow\infty}K_n(d)$ numerically, we generalise
an argument due to \cite{chen}: For $n$ sufficiently
large, the sums $K_n(d)$ for different values of
$n$ will all intersect with each other at approximately
the same point $d=D_H$. It provides a method
to estimate the Hausdorff dimension.

In Fig. 2 we show $\ln K_5(d)$, $\ln K_6(d)$, $\ln K_7(d)$
and $\ln K_8(d)$ for $R=0.37$. The curves in fact intersect
approximately at the same point giving $D_H(0.37)=0.87\pm 0.01$.
In Fig. 3 (boxes) we show a graph of $D_H$ computed by this
method for $0.37\leq R\leq 0.48$ (it includes
regions with $D_H>1$ and with $D_H\leq 1$).
In the same figure (stars) we also show the {\it box-counting
dimension} $D_c$ of $\cap_{m=1}^{\infty}C_m$.
$D_c$ was calculated by approximating $\cap_{m=1}^{\infty}C_m$
by $C_n$ for a sufficiently large $n$.
The number of squares $N(\varepsilon)$
needed to cover $C_n$ was counted for different values of the
edge length $\varepsilon$, and the dimension $D_c$
was then obtained by fitting $\ln N(\varepsilon)=
-D_c \ln \varepsilon$.
The comparison between $D_c$ and $D_H$ in Fig. 3
suggests that $D_c(R)=D_H(R)$.

Some remarks about the computation of the box-counting
dimension have to be made. The accurate evaluation of
$D_c$ requires the use of relatively small values
for $\varepsilon$. On the other hand, $C_n$ has to
reflect the fractal property of $\cap_{m=1}^{\infty}C_m$,
condition satisfied if $|c_{n,j}|<\varepsilon$ for all $j$.
It implies the need of large values for $n$ (we used $n=15$
in our computations), demanding great computational
efforts in finding $C_n$ when $D_c$ is less than one.
The remedy is based on the use of a modified
PIM (proper-interior-maximum) procedure\footnote{
This technique can also be employed in the
computation of $|c_{n,j}|$, to calculate $D_H$.
}
\cite{nuss}.
In the first step, we start with a sample of initial
conditions on $P$ taken at the corners of a grid.
From this set we store those
points that are in $C_1$ as well as their first
neighbours. We denote the stored set by $\bar{C}_1$.
In the second step, we reduce the edge length
of the grid and iterate only those points of the
new grid that are in the region covered by $\bar{C}_1$.
Then we store the points that are in $C_2$
together with their first neighbours. We denote
this stored set by $\bar{C}_2$, and so on.
After $n$ steps we obtain the desired approximation
for $C_n$.

Back to Fig. 1, in the centre of the figure we
see an approximately self-similar structure.
Each step of the construction of the Cantor set divides
one disc in three smaller deformed discs.
If we assume that the
self-similarity is exact, we can introduce
$\lambda_1$, $\lambda_2$ and $\lambda_3$,
the linear scales by which the discs are
reduced in one step. Under such hypothesis
we have that Hausdorff and box-counting dimensions
are equal and satisfy \cite{sandau}
\begin{equation}
\lambda_1^D +\lambda_2^D +\lambda_3^D =1.
\label{16}
\end{equation}
Even through the self-similarity is not exact in our
case, (\ref{16}) is expected to provide a good estimate
for $D_H$ (and $D_c$).
Fig. 1 also shows that the factors by which the discs
are reduced are approximately the same.
It allows us to take
$\lambda_1=\lambda_2=\lambda_3\equiv\lambda$,
and we refer to the resulting dimension
\begin{equation}
D_s=-\frac{\ln 3}{\ln \lambda }
\label{17}
\end{equation}
as the {\it self-similarity dimension}.
The computation of $\lambda$ can be performed
by a Monte Carlo method.
We determine the areas of $C_n$ and $C_{n+1}$,
for a sufficiently large $n$, by evolving
a sample of random initial conditions.
 An estimate of $\lambda $ is then
given by
\begin{equation}
\lambda_n =\sqrt{\frac{1}{3}\times
\frac{\mbox{area}(C_{n+1})}{\mbox{area}(C_n)}}\;\; .
\label{18}
\end{equation}
A better statistics is obtained by taking
$\lambda$ as the average on 
$\lambda_n, \lambda_{n+1}, ...,\lambda_{n+k}$
for an adequate $k$.
For the tetrahedron scatterer, the 
value of $\lambda$ is very stable in relation to
$n$ and $k$, what allows us to compute
a well defined $\lambda$
with not much computational effort.
The resulting dimension is shown in Fig. 3 (triangles)
for $0.36\leq R\leq 0.49$
and is consistent with $D_H$ and $D_c$ computed
directly from their definitions.
We believe that $D_s$ can be used to estimate the common
value of $D_H$ and $D_c$ in most systems where some kind of
statistical self-similarity takes place. A rigorous statement
in this direction requires, however, further investigations.

The dimension ($D$) that we computed corresponds to the
intersection of the stable manifold of
the invariant set (repellor) with the plane $P$.
In the 5-dimensional energy surface,
the dimension of the stable manifold will be $d_s=3+D$.
The time reversibility of the dynamics implies that the
dimension of the unstable manifold $d_u$
will be equal to $d_s$.
Since the invariant set is the intersection of the stable
and unstable manifolds, its dimension will be
$d_i=d_s+d_u-5=2D+1$. Therefore, the dimensions
$d_s$, $d_u$ and $d_i$ can be calculated by
computing $D$ with the techniques employed here.
In this procedure, the critical value of the
radius ($R_c$) where $D=1$ ($R_c\approx 0.41$,
see Fig. 3) does not play any particular role.
In this context, to claim that the dynamics
of the scattering systems
is chaotic for $D>1$ ($d_i>3$) and regular for 
$D\leq 1$ ($d_i\leq 3$) sounds largely a matter
of semantics. What is sure is that there is a
chaotic set of trapped orbits whose dimension
increases with the radius $R$.
The presence of this chaotic set prevents the
existence of integrals of motion besides the energy
(the system is non-integrable).

In the computation of $D_H$, $D_c$ and $D_s$,
we explored the hierarchical structure of levels of the
time delay function in order to infer the dimension of
its set of singularities ($\cap_{m=1}^{\infty}C_m$).
The dimension of the singularities of the time delay
seems to be the most
fundamental quantity that we can measure to characterise
invariant sets of scattering systems. The result
involves no ambiguity since it will lead unequivocally to the
dimension of the future invariant set. The same is not
true for basin boundaries, for instance.
See Fig. 4, where we show the basins of the tetrahedron
scatterer defined by particles scattered to
$x\rightarrow +\infty$ and particles scattered to
$x\rightarrow -\infty$, for $R=0.48$ (a) and $R=0.37$ (b).
The basin boundaries present smooth 1-dimensional
parts on arbitrarily fine scales.
These smooth parts are not on the future invariant set
and, although immaterial when $R=0.48$,
they do affect the basin boundary dimension
when $R=0.37$. In this case the dimension of the
fractal part ($I_f\cap P$) is less than the
dimension of the smooth parts. Smooth parts also
forbid the use of uncertainty methods to estimate the
dimension of the future invariant set
from the time delay function.
In computing the uncertainty dimension of the
singularities, the uncertainty method effectively
computes the box-counting dimension of the set of all
discontinuities. The time delay is discontinuous on the
(1-dimensional) frontiers of all $C_n$ (see Fig. 1),
what implies that the
result will be always greater than or equal to 1.
The same happens with scattering functions.

Finally, we remark that methods to directly estimate
the Hausdorff dimension of {\it attractor} sets
(taking into account the infimum in (\ref{1}))
was already discussed (see \cite{haus} and references
therein). The main difficulty in adapting these methods
to the case of {\it nonattracting} sets is that they
require a sample of points on the set we want to measure.
Even though a simple matter in the case of attractors,
the computation of these points is nontrivial for
nonattracting sets. A possibility would be the 
use of the PIM-triple procedure \cite{nuss}
to find trajectories which stay near the
invariant set for arbitrarily long periods of time.
The resulting methods are, however, far more complicated
than those introduced here.

In summary, we have presented, through an example,
methods to estimate the Hausdorff dimension of
nonattracting invariant sets of scattering systems.
The methods apply, albeit not only, to systems
with invariant set dimension so smaller that the
uncertainty methods do not work. It includes cases
where the codimension of the future invariant set is
approximately equal to or greater than one.
We stress that these
methods are general and can be used in phase spaces of
any dimension. And since they allow, in principle,
the computation of arbitrarily small dimensions,
these techniques should be useful to study
how chaotic scattering comes about as a system
parameter is varied \cite{bog}.
In particular, it may be interesting in investigations
about routes to chaos in three dimensional scattering. 

The authors thank Fapesp and CNPq for financial support.

\newpage

\begin{figure}
\caption{$C_n$ for $R=0.37$:
$C_1$ (light grey), $C_2$ (grey), $C_3$ (dark grey) and $C_4$ (black). 
}
\label{fig1}
\end{figure}

\begin{figure}
\caption{$\ln K_n$ for $R=0.37$:
$n=5$ (least inclined curve), $n=6$, $n=7$ and $n=8$
(most inclined curve). 
}
\label{fig2}
\end{figure}

\begin{figure}
\caption{Estimates of the dimension of
$\cap_{m=1}^{\infty}C_m$
as a function of $R$: $D_H$ (boxes),
$D_c$ (stars) and $D_s$ (triangles).
The size of the boxes, stars and triangles corresponds
approximately to the statistical uncertainty of
the dimension. 
}
\label{fig3}
\end{figure}

\begin{figure}
\caption{Portrait of the basins of the tetrahedron
scatterer for: (a) $R=0.48$; (b) $R=0.37$.
The initial conditions were chosen on a grid of
$400\times 400$, on the plane $P$. Regions in black
and white correspond to orbits that escape to
$x\rightarrow +\infty $ and $x\rightarrow -\infty $,
respectively. Regions in grey correspond to orbits
not scattered.
}
\label{fig4}
\end{figure}

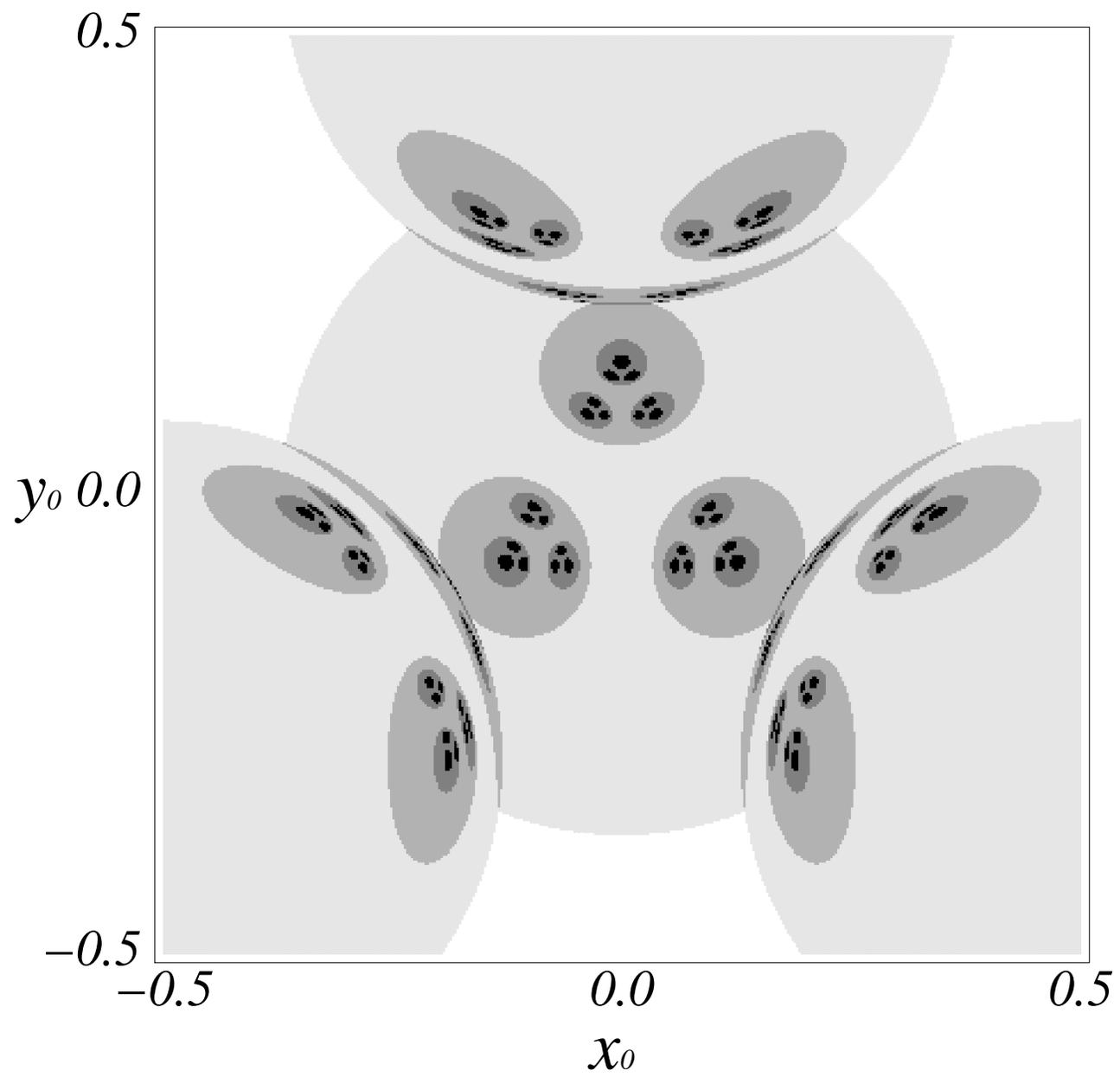
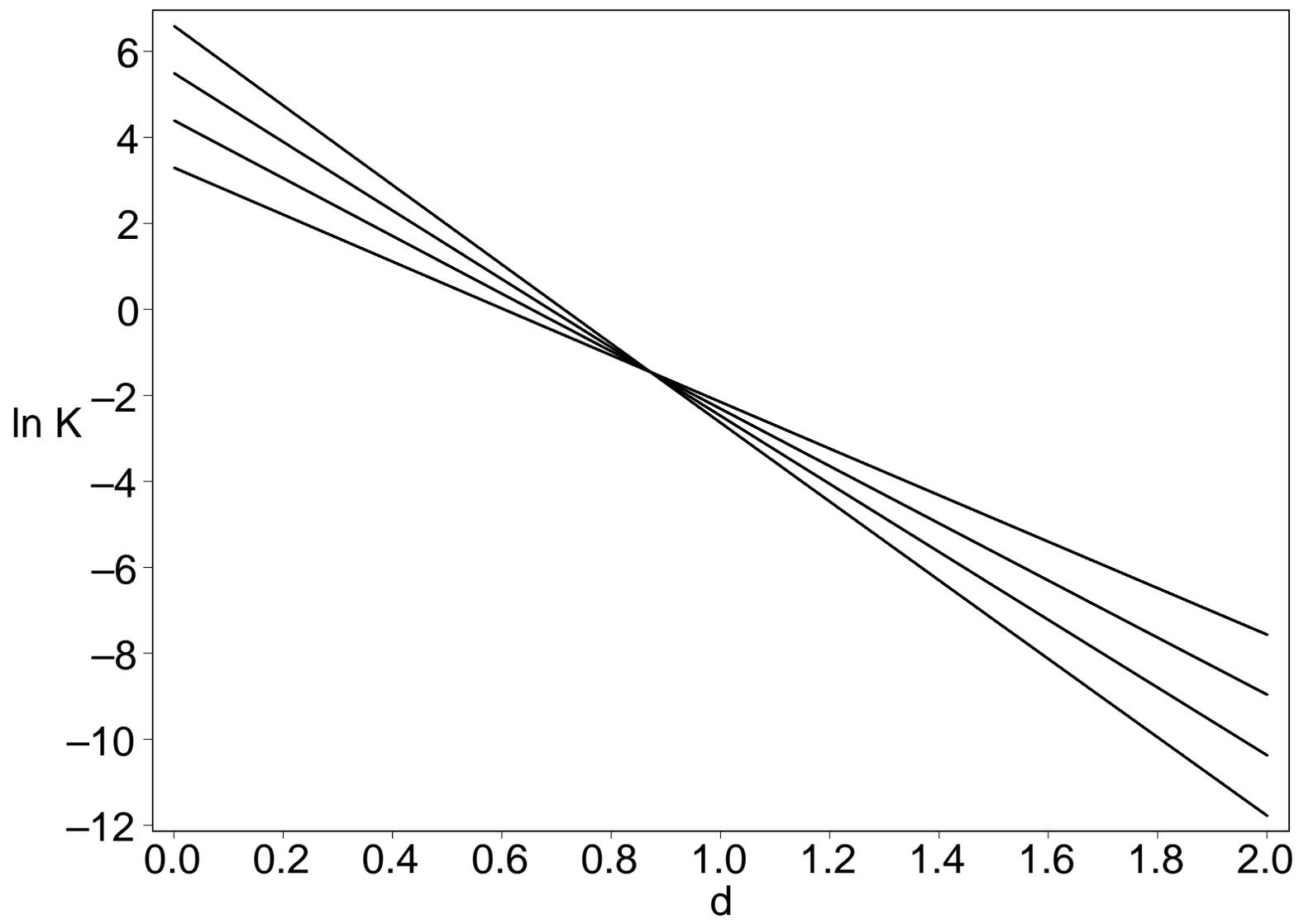
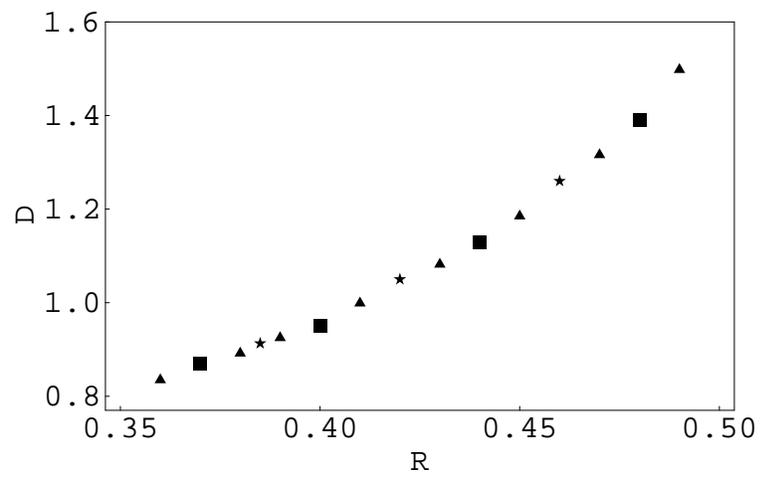
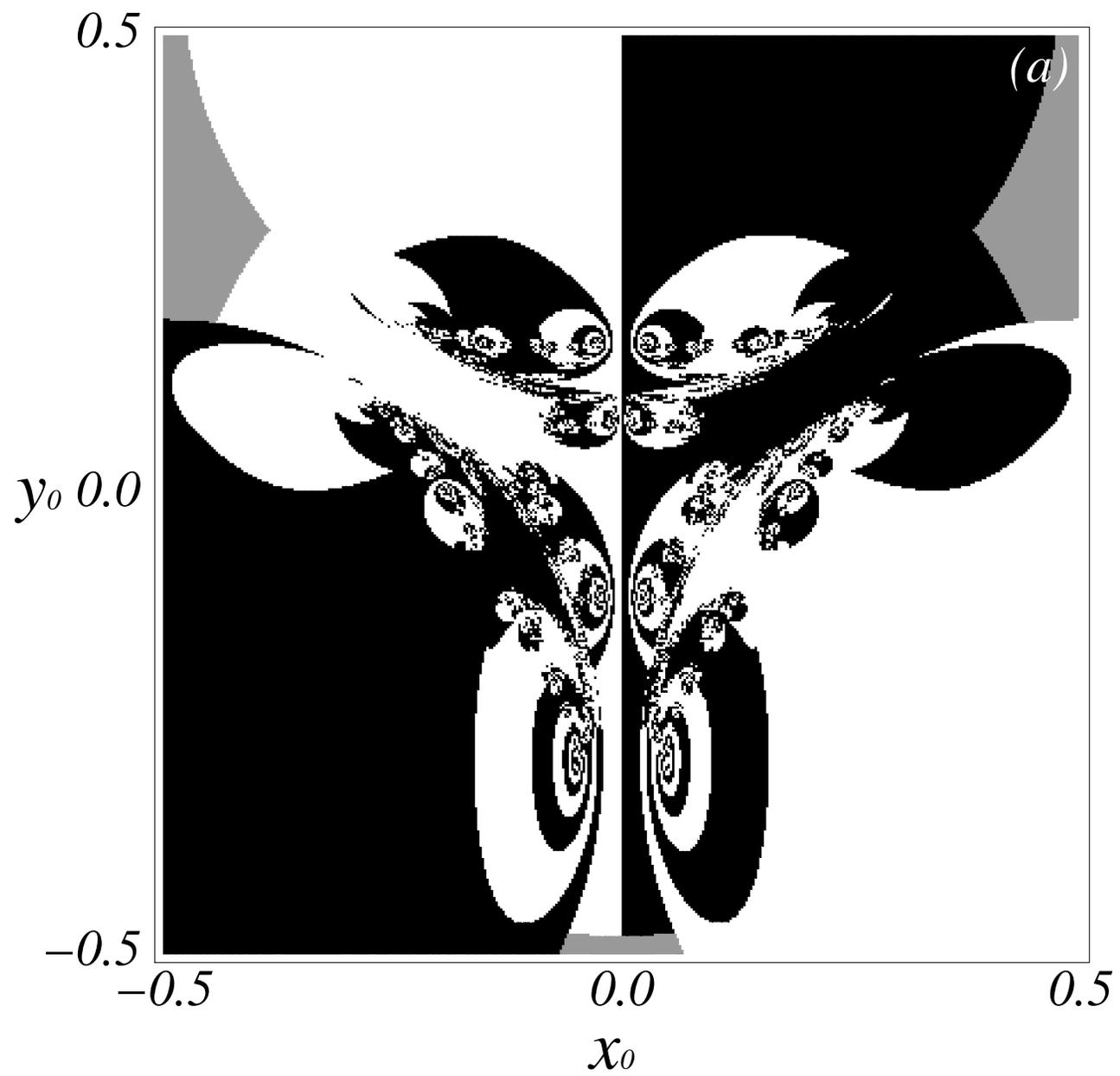
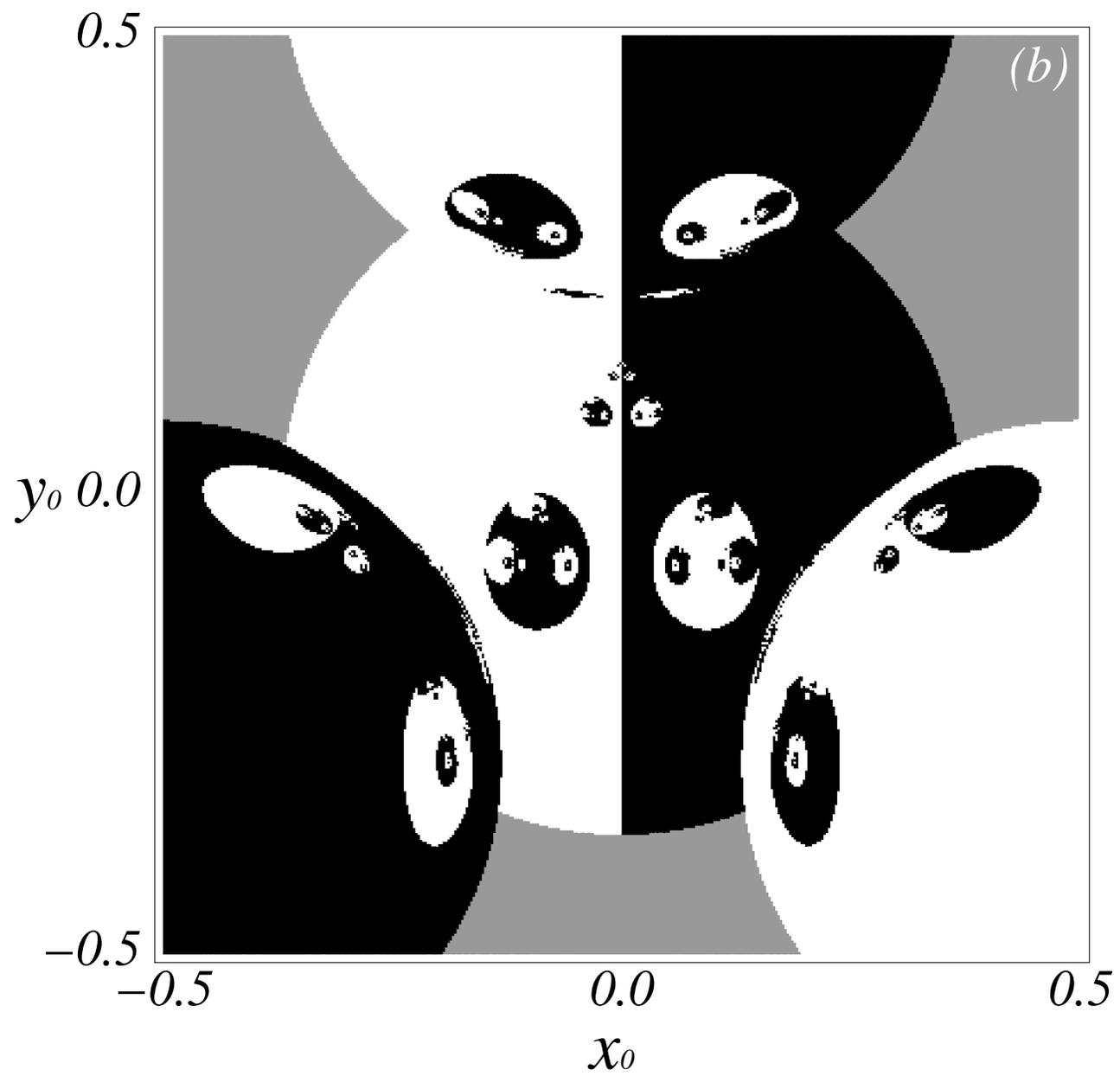
\end{document}